\documentclass{article}
\usepackage[pdftex]{graphicx}
\usepackage{amsmath}
\usepackage{amsopn}
\usepackage{amsfonts}
\usepackage{amssymb}
\usepackage[square]{natbib}
\usepackage{url}
\usepackage{subfigure}
\usepackage{booktabs}
\usepackage{rotating}
\usepackage[hang,footnotesize,bf]{caption}
\usepackage{setspace}
\usepackage{multirow}
\usepackage[]{algorithm}
\usepackage{algorithmic}
\usepackage{epstopdf}

 \usepackage{lineno}
 \linenumbers*[1]

\usepackage{colortbl}
\arrayrulecolor{black}
\doublerulesepcolor{black}

\newcommand*{\IsDraft}{false}

\newcommand*{\halfwidth}{8cm}

\newcommand{\be}{\begin{equation}}
\newcommand{\ee}{\end{equation}}
\newcommand{\ba}{\begin{eqnarray}}
\newcommand{\ea}{\end{eqnarray}}

\begin{document}
\title{Adaptively Smoothed Seismicity Earthquake Forecasts for Italy}

\author{
Maximilian J.\ Werner\textsuperscript{1}\textsuperscript{*},
Agn\`{e}s Helmstetter\textsuperscript{2},
David D. Jackson\textsuperscript{3},\\
Yan Y. Kagan\textsuperscript{3},
and Stefan Wiemer\textsuperscript{1}}
\date{\today}
\maketitle

\begin{center}
first version submitted 23 March 2010 to the CSEP-Italy special issue of the Annals of Geophysics \\
revised version submitted 28 June 2010\\
\end{center}

\vskip0.3cm

\noindent \textsuperscript{1} Swiss Seismological Service, Institute of Geophysics, ETH Zurich, Switzerland. \\
\textsuperscript{2} Laboratoire de G\'eophysique Interne et Tectonophysique, Universit\'e Joseph Fourier and Centre National de la Recherche Scientifique, Grenoble, France. \\
\textsuperscript{3} Department of Earth and Space Sciences, University of California, Los Angeles, USA. \\
\noindent \textsuperscript{*} Corresponding author:  \\
\hspace* {5cm} Maximilian J. Werner \\
\hspace* {5cm} Swiss Seismological Service \\
\hspace* {5cm} Institute of Geophysics\\
\hspace* {5cm} ETH Zurich\\
\hspace* {5cm} Sonneggstr. 5\\
\hspace* {5cm} 8092 Zurich, Switzerland\\
\hspace* {5cm} {\bf mwerner@sed.ethz.ch}

\doublespacing

\abstract

We present a model for estimating the probabilities of future earthquakes of magnitudes $m \geq 4.95$ in Italy. The model, a slightly modified version of the one proposed for California by \citet{Helmstetter-et-al2007} and \citet{Werner-et-al2009b}, approximates seismicity by a spatially heterogeneous, temporally homogeneous Poisson point process. The temporal, spatial and magnitude dimensions are entirely decoupled. Magnitudes are independently and identically distributed according to a tapered Gutenberg-Richter magnitude distribution. We estimated the spatial distribution of future seismicity by smoothing the locations of past earthquakes listed in two Italian catalogs: a short instrumental catalog and a longer instrumental and historical catalog. The bandwidth of the adaptive spatial kernel is estimated by optimizing the predictive power of the kernel estimate of the spatial earthquake density in retrospective forecasts. When available and trustworthy, we used small earthquakes $m\geq2.95$ to illuminate active fault structures and likely future epicenters. By calibrating the model on two catalogs of different duration to create two forecasts, we intend to quantify the loss (or gain) of predictability incurred when only a short but recent data record is available. Both forecasts, scaled to five and ten years, were submitted to the Italian prospective forecasting experiment of the global Collaboratory for the Study of Earthquake Predictability (CSEP). An earlier forecast from the model was submitted by \citet{Helmstetter-et-al2007} to the Regional Earthquake Likelihood Model (RELM) experiment in California, and, with over half of the five-year experiment over, the forecast performs better than its competitors. 

\section{Introduction}

In this article, we document the calibration of a previously published, time-independent model of earthquake occurrences to the region of Italy. We extracted probabilities of future $m \geq 4.95$ shocks  for a five- and ten-year period in a format suitable for prospective testing within the Italian earthquake predictability experiment \citep{Schorlemmer-et-al2010}. Previously, \citet{Helmstetter-et-al2007} calculated a probabilistic earthquake forecast for $m \geq 4.95$ for the region of California over a five year duration. The forecast is currently being tested within the Regional Earthquake Likelihood Model (RELM) experiment \citep{Field2007}. After more than half of the five years over, the forecast cannot be rejected by a suite of tests and performs better than competing forecasts \citep[][]{Schorlemmer-et-al2010r}. \citet{Werner-et-al2009b} made small modifications to the model by \citet{Helmstetter-et-al2007} and re-calibrated it on updated data to generate a new earthquake forecast for California. This forecast is under test within the California branch of the global Collaboratory for the Study of Earthquake Predictability (CSEP) \citep{Jordan2006, Zechar-et-al2009}. To calculate future earthquake potential in Italy, we used the same model with some minor modifications. One modification concerns the estimation of the completeness threshold, which was difficult to estimate at the small spatial scales that were possible with the high quality data set available in California  \citep{Werner-et-al2009b, Helmstetter-et-al2007}. Instead, we set a single magnitude threshold for the entire region. 

Smoothed seismicity models, such as the present one, usually do not incorporate geological or tectonic observations. Rather, the models are calibrated on the seismicity data available from earthquake catalogs. One may justifiably question the hypothesized validity that the short (decadal) periods covered by high-quality instrumental catalogs suffice to forecast the locations of future large earthquakes that have very low occurrence probabilities. Even if the spatial distribution of seismicity is reasonably stable up to geological timescales, estimating this distribution from a short time window of observations is difficult. 

A partial solution to this problem is to estimate a predictive spatial distribution, rather than the observed spatial distribution. That is, rather than estimating the density of past earthquakes, one divides available data into separate learning and target sets to estimate a predictive density from the learning catalog that is evaluated and optimized on the target quakes. This cross-validation method we employed generates smoother forecasts than a simple kernel density estimation method because the locations of future -- rather than past -- earthquakes are predicted, and such locations might be in regions of little previous seismicity. Nonetheless, this is only a partial solution: \citet{KaganJackson1994} conjectured that the optimal predictive horizon of a forecast based on smoothed seismicity scales proportionally with the learning catalog because of spatio-temporal clustering. Thus, the forecasts by \citep{Werner-et-al2009b} and \citet{Helmstetter-et-al2007}, which were calculated from about 25 years of high quality Californian data, should perform well for moderate earthquakes over similar timescales, but longer periods relevant for seismic hazard estimates and building codes might require longer input data sets. On the other hand, according to  \citet{KaganJackson1994}, it remains an untested hypothesis that seismicity estimates based on geological observations, i.e. the earthquake history of several thousands of years, can provide more predictive and relevant information for engineering design than high-quality low-threshold instrumental catalogs. Their argument is based on two points: first, the quality of geologic observations for seismic hazard is often low compared to the quality of modern earthquake catalogs; second, seismicity exhibits long-term spatio-temporal variations that necessitate an appropriate weighting of the information contained in observations of recent earthquakes and those that occurred in the distant past.

To begin to quantify the impact of the duration of the learning catalog on the predictive power of smoothed seismicity forecasts, we calculated two forecasts of the smoothed seismicity model by calibrating the model on two catalogs of different duration. One forecast is based on the most recent thirty years of instrumental data, while the other one is calculated from a century of combined historic and instrumental data. We submitted the forecasts to the European CSEP Testing Center at ETH Zurich, to be tested and compared with competing forecasts within the framework of the Italian earthquake predictability experiment CSEP-Italy \citep[see][]{Schorlemmer-et-al2010}. Comparing the two forecasts' performance might shed light on the impact of the length of the learning catalog.

The article is structured as follows. In section \ref{sec:data}, we describe the Italian earthquake catalogs from which we estimated future earthquake potential in Italy. Section \ref{sec:model} describes the model and its calibration on the two data sets. We present the earthquake forecasts in section \ref{sec:results} before concluding in section \ref{sec:conc}.

\section{Data}
\label{sec:data}

\subsection{The CSI 1.1 Catalog 1981--2002}
\label{sec:CSI}

For more details about the catalogs discussed here and below, see \citep{Schorlemmer-et-al2010} and references therein. As primary data source for the forecast based on smoothed instrumental seismicity, we used the Catalogo della Sismicit\`a Italiana (catalog of Italian seismicity, CSI 1.1) \citep{Castello-et-al2007, Chiarabba-et-al2005} available from \url{http://www.cseptesting.org/regions/italy}. The CSI catalog spans the time period from 1981 until 2002 and reports local magnitudes, in agreement with the magnitude scale that will be used during the prospective evaluation of forecasts. \citet{Schorlemmer-et-al2010} found a clear change in earthquake numbers per year in 1984 due to the numerous network changes in the early 1980s and recommend using the CSI data from mid 1984 onwards. We therefore selected earthquakes listed in the CSI catalog from 1 July 1984 until 31 December 2002 within the CSEP-Italy collection region defined by \citet{Schorlemmer-et-al2010}. To avoid possible contamination from quarry blasts and volcanic microseismicity, we set a uniform completeness magnitude threshold of $m_t=2.95$ for the entire collection region, which is higher than the threshold of $m_t=2.5$ calculated by \citet{Schorlemmer-et-al2010} for onshore seismicity.

\subsection{The BSI Catalog 2003--2009}

For prospective tests of the submitted forecasts \citep[see][]{Schorlemmer-et-al2010}, the Bollettino Sismico Italiano (BSI) earthquake catalog recorded by the Istituto Nazionale di Geofisica e Vulcanologia (INGV) \citep{BSI2002, Amato-et-al2006} will be used. The BSI is available at \url{http://bollettinosismico.rm.ingv.it}, and since July 2007 at \url{http://ISIDe.rm.ingv.it/}. We used earthquakes listed in the BSI from 1 January 2003 until 25 June 2009. Because the data quality of small quakes in the BSI catalog between 2003 and 2005 is questionable, we applied a relatively high magnitude threshold of $m_t=2.95$. For the forecast based on recent instrumental observations, we merged the BSI and CSI catalogs (hereinafter called the ``merged instrumental catalog" or MIC).

\subsection{The CPTI08 Catalog 1901--2006}
\label{sec:CPTI}

For the forecast based on instrumental and historical seismicity, we used the Catalogo Parametrico dei Terremoti Italiani (parametric catalog of Italian earthquakes, CPTI08) \citep{CPTI08} available from \url{http://www.cseptesting.org/regions/italy}. The CPTI08 catalog, a preliminary revision of the 2004 CPTI04 catalog \citep{CPTI2004}, covers the period from 1901 until 2006 and is based on both instrumental and historical observations. The catalog lists moment magnitudes that were estimated either from macroseismic data or calculated using linear (``ad-hoc" \citep{CPTI08}) regression relations between surface wave, body wave or local magnitudes \citep{MPS2004}. \citet{Castellaro-et-al2006} showed that standard linear regression (SLR) can lead to biased and uncertain magnitude conversions because magnitude observations and associated errors violate the simplifying assumptions of SLR. As a remedy, \citet{Castellaro-et-al2006} advocate the use of general orthogonal regression. Therefore, the CPTI08 catalog can be expected to contain biases and large uncertainties (like most historical catalogs) that might affect the quality of earthquake forecasts. In their global forecasts, \citet{KaganJackson2010a} therefore used more homogeneous catalogs: the global Centroid Moment Tensor (CMT) catalog \citep{Ekstrom-et-al2005} and the global Preliminary Determination of Epicenters (PDE) catalog by the U.S. Geological Survey \citep{PDE2001}. Nonetheless, the CPTI08 catalog offers a much longer training catalog (a century) than either of these two global catalogs (30 to 40 years), allowing us to investigate the effect of the length of the training catalog on the generated forecasts. Therefore, we accepted the potential shortcomings of the CPTI08 catalog for this study. Because the prospective experiment will use local magnitudes, we converted the moment magnitudes to local magnitudes using the same regression equation that was used to convert the original local magnitudes to moment magnitudes for the creation of the CPTI catalog \citep{MPS2004, Schorlemmer-et-al2010}:
\begin{equation}
m_L=1.231 (m_W - 1.145) \ .
\label{eq:ML}
\end{equation}

As for the BSI and CSI catalogs, we only selected shocks within the collection region and with depths shallower than 30km. Some quakes, mostly during the early part of the CPTI catalog, were not assigned depths. We included these earthquakes as observations within the testing region because it is very unlikely that they were deeper than 30km \citep[see also][]{Schorlemmer-et-al2010}. We selected earthquakes with moment magnitudes $m_W \geq 4.75$ \citep{Schorlemmer-et-al2010}, which corresponds to local magnitudes $m_L \geq 4.45$.

\section{Model Calibration}
\label{sec:model}

The model has previously been documented by \citet{Helmstetter-et-al2007} and \citet{Werner-et-al2009b}, and we will only provide a brief overview here. First, earthquake catalogs were declustered to remove the strong influence of triggered sequences (section \ref{sec:decl}); if we did not decluster, we would need to use a more complicated, time-dependent model that removes the influence of aftershocks with the Omori-Utsu law \citep{Omori1894, Utsu-et-al1995}. Once declustered, the seismicity was smoothed with an adaptive power-law kernel (section \ref{sec:ass}). The bandwidth of the kernel at each earthquake epicenter adapts to the distance to the $k$th nearest neighbor. To estimate the optimal number of neighbors to include in the smoothing, we divided the catalog into two non-overlapping sets: a learning catalog and a testing catalog. In section \ref{sec:smoothopt}, we determine the optimal number of neighbors by calculating the spatial density of seismicity from the learning catalog and evaluating its predictive power on the testing catalog. The spatial density was scaled to the number of expected earthquakes by using the mean number of observed earthquakes (section \ref{sec:prednum}). Finally, to obtain a rate-space-magnitude forecast, we multiplied the scaled spatial density by a tapered Gutenberg-Richter magnitude-frequency distribution (section \ref{sec:md}). In contrast to the model presented by \citet{Helmstetter-et-al2007} and \citet{Werner-et-al2009b}, we used a homogeneous threshold for the completeness magnitude because the previous method, which estimated the threshold as a function of space, did not perform well for Italy.

\subsection{Declustering}
\label{sec:decl}
To estimate the spatial distribution of spontaneous earthquakes, we used the declustering algorithm proposed by \citet{Reasenberg1985}, modified slightly by \citet{Helmstetter-et-al2007} and \citet{Werner-et-al2009b}. As in these prior studies, we set the input parameters to $r_{fact} = 8$, $x_k=0.5$, $p_1=0.95$, $\tau_{min}=1$ day and $\tau_{max}= 5$ days. We varied $x_{meff}$ according to the different learning catalogs we used. As the interaction distance, we used the scaling $r=0.01 \times 10^{0.5M}$ km suggested by \citet{WellsCoppersmith1994} instead of $r=0.011 \times 10^{0.4M}$ km and $r < 30$ km in Reasenberg's algorithm. Like  \citet{Werner-et-al2009b}, we set the localization errors to $1$ km horizontal and $2$ km vertical. We found that about $80\%$ of earthquakes in the merged instrumental catalog (MIC) are spontaneous, while in the CPTI catalog, about $92\%$  of all shocks are independent according to Reasenberg's classification. 

\subsection{Adaptive Kernel Smoothing of Declustered Seismicity}
\label{sec:ass}
We estimated the density of spontaneous seismicity in each $0.1$ by $0.1$ degree cell by smoothing the location of each earthquake $i$ with an isotropic adaptive power-law kernel $K_{d_i} (\vec{r})$:
\be
K_{d_i} \left(\vec{r}\right) = \frac{C(d_i)}{\left(|\vec{r}|^2+d_i^2\right)^{1.5}}
\label{eq:Kpl}
\ee
where $d_i$ is the adaptive smoothing distance and $C(d_i)$ is a normalizing factor, so that the integral of $K_{d_i} \left(\vec{r}\right)$ over an infinite area equals 1.

We measured the smoothing distance $d_i$ associated with an earthquake $i$ as the horizontal distance between event $i$ and its $k$th closest neighbor. The number of neighbors, $k$, is an adjustable parameter, estimated by optimizing the forecasts (see section \ref{sec:smoothopt}). We imposed $d_i>0.5$ km to account for location uncertainty. The kernel bandwith $d_i$ thus decreases if the density of seismicity is high at the location $\vec{r_i}$ of the earthquake $i$, so that we have higher resolution (smaller $d_i$) where the density is higher. 

The density at any point $\vec{r}$ was estimated by
\be
\mu(\vec{r}) = \sum_{i=1}^{N_l} K_{d_i} \left(\vec{r}-\vec{r_i} \right)
\label{eq:mu}
\ee
where $N_l$ is the total number of earthquakes in the learning catalog. However, the forecasts are given as an expected number of events in each cell of $0.1^\circ$. We therefore integrated equation (\ref{eq:Kpl}) over each cell and summed over all contributing earthquakes to obtain the seismicity rate of each cell. 

\subsection{Optimizing the Spatial Smoothing}
\label{sec:smoothopt}

We estimated the parameter $k$, the number of neighbors used to compute the smoothing distance $d_i$ in equation (\ref{eq:mu}), by maximizing the likelihood of the model. We built the model $\mu'(i_x,i_y)$ in each cell $(i_x,i_y)$ from the data in the learning catalog and evaluated the likelihood of target earthquakes in the testing catalog. Because we assumed independence of the spatial density from the magnitude distribution and the total expected number of events, we optimized the normalized spatial density estimate in each cell $(i_x,i_y)$ using
\be
\mu^*(i_x,i_y) = \frac{\mu'(i_x,i_y) N_t}{\sum_{i_x} \sum_{i_y}\mu'(i_x,i_y)}
\label{eq:munorm}
\ee
where $N_t$ is the number of observed target events. The expected number of events for the model $\mu^*$ thus equals the observed number $N_t$. 

The log-likelihood of the model is given by
\be
L = \sum_{i_x} \sum_{i_y} \log p \left[ \mu^* (i_x,i_y), n \right]
\label{eq:L}
\ee
where $n$ is the number of events that occurred in cell $(i_x,i_y)$. To adhere to the rules of the CSEP-Italy predictability experiment, we assumed that the probability $p$ of observing $n$ events in cell $(i_x,i_y)$ given a forecast of $\mu^* (i_x,i_y)$ in that cell is given by the Poisson distribution
\be
p \left[ \mu^* (i_x,i_y), n \right] = \left[ \mu^*(i_x,i_y)  \right]^n \frac{\exp \left[ -\mu^*(i_x,i_y) \right]}{n!}
\label{eq:Poi}
\ee
We built a large set of background models $\mu^*$ by varying (i) the starting times, end times and magnitude thresholds of the learning and testing catalogs, and (ii) the catalog (either the MIC or the CPTI catalog). We evaluated the performance of each model by calculating its probability gain per target earthquake relative to a model with a uniform spatial density:
\be
G = \exp \left( \frac{L - L_{0}}{N_t} \right)
\label{eq:G}
\ee
where $L_{0}$ is the log-likelihood of the spatially homogeneous model. 

\subsection{Results of the Spatial Optimization}
\label{sec:resultsopt}

Tables \ref{tab:merged} and \ref{tab:cpti} show the results of the spatial optimization on the MIC and the CPTI catalog, respectively. For each model, we found the optimal smoothing parameter $k$ in the range $[1,50]$ by choosing the value that maximized the likelihood of the target earthquakes in the target catalog given the smoothed spatial density estimated from the learning catalog. We varied the magnitude threshold of the input catalog to test whether including small earthquakes results in greater predictability of future $m \geq 4.95$ earthquakes. We also changed the target periods to test the robustness of the results. 

In Figure \ref{fig:G}, we show the probability gains per earthquake against the magnitude threshold of the two learning catalogs. For comparison, we also included the gains obtained for the five-year period 2004-2008 (inclusive) in California by \citet{Werner-et-al2009b}. The gains obtained in Italy fluctuate strongly for different target periods, and it is difficult to detect a systematic trend in the gain as a function of the magnitude threshold of the learning catalog. In contrast to California, where around 25 earthquakes $m \geq 4.95$ tend to occur per five years, Italy experiences far fewer shocks of equal size; during the 1992-1996 period in the CPTI catalog, gains were calculated from only two target earthquakes. The small sample size of target earthquakes might explain the observed fluctuations in the calculated gains (see also below and section \ref{sec:conc}). 

For the target period 1994-1998 of the MIC, the gains are especially small for low thresholds. For $m_t=2.95$, the smoothing parameter reached the maximum value $50$ (model 7 in Table \ref{tab:merged}), realizing a gain smaller than unity, i.e. the uniform forecast outperforms the smoothed seismicity forecast (further increasing the amount of smoothing eventually leads to a uniform density, such that the gain would equal unity). The long-range smoothing required by the target events can be traced back to the occurrence of  three earthquakes in 2002:  the 6 September Sicily earthquake north-east of Palermo, the 31 October Molise earthquake and one of its aftershocks. The predicted densities in the relevant cells is increased by a factor of almost ten as the algorithm increases the smoothing from $k=1$ to $k=50$ (see also the discussion in section \ref{sec:conc}). 

Whenever target earthquakes occur in previously active regions, the optimal amount of smoothing is small ($k=1$), and the gains tend to be higher (see, e.g., model 13 in Table \ref{tab:merged} for the 1994-1998 target period that includes the 1997 Colfiorito earthquake sequence). However, exceptions exist to the expected anti-correlation between $k$ and $G$: the 15 target earthquakes during the 1997-2001 target period of the CPTI catalog were forecast best with a smoothing parameter $k=15$ yet realize a gain per earthquake of $G=3.41$.

To calculate the spatial densities for the final forecasts for the predictability experiment (model 20$^{!}$ in Table \ref{tab:merged} obtained from the MIC and model 14$^{!}$ in Table \ref{tab:cpti} obtained from the CPTI catalog), we had to decide which magnitude threshold to apply to the learning catalog, which smoothing parameter to use, and whether to use all existing data up until the end of the two catalogs. We decided to use all available data in each catalog for the final density estimate so that the forecasts could benefit from as much data as possible. Moreover, despite the observed variability in gains against the magnitude threshold of the input catalog (discussed above), Figure \ref{fig:G} shows that, on average, there seems to be an advantage in including small earthquakes for estimating the predictive spatial density (see also the discussion in section \ref{sec:conc}). Therefore, to calculate the spatial densities for the final forecasts, we used the lowest reliable magnitude threshold for each catalog. Lastly, we decided to use an optimal smoothing parameter of $k=6$, despite the large variability across magnitude thresholds and target periods, because the resulting density is slightly smoother than the one obtained from the median ($k=5$) of the optimal values for the lowest magnitude thresholds and because both \citet{Werner-et-al2009b} and \citet{Helmstetter-et-al2007} used the same value. The two final predictive spatial densities based on the MIC and the CPTI catalog are model 20$^{!}$ in Table \ref{tab:merged} and model 14$^{!}$ in Table \ref{tab:cpti}, respectively. We discuss future improvements of the spatial optimization method in section \ref{sec:conc}.

\subsection{Magnitude Distribution}
\label{sec:md}

We assumed that the cumulative magnitude probability distribution follows a tapered Gutenberg-Richter magnitude frequency distribution \citep{GutenbergRichter1944} with a uniform b-value and corner magnitude $m_c$ \citep[][Eq. (10)]{Helmstetter-et-al2007}
\begin{equation}
P(m)=10^{-b(m-m_{min})} \exp\left[ 10^{1.5(m_{min}-m_c)}-10^{1.5(m-m_c)}\right]
\label{tGR}
\end{equation}
with a minimum target magnitude $m_{min}=4.95$ (for the five- and ten-year CSEP forecast group). \citet[][p. 2393]{BirdKagan2004} classified the tectonic setting of Italy as an orogen situated at a continental convergent boundary and estimated $m_c = 8.46^{+0.21}_{-0.39}$. \citet[][Figure 1, Table 1]{Kagan-et-al2010} assigned onshore Italy to the category of ``active continent" with $m_c=7.59^{+0.72}_{-0.25}$ and Italy's southern off-shore region to ``trench" with $m_c=8.57^{+?}_{-0.35}$. For simplicity, we set a uniform value of $m_c=8.0$ to reflect these studies. This is likely to be a conservative choice for the most seismically active region of onshore Italy. We further used a b-value equal to one (a maximum likelihood estimate based on magnitudes above $m_t=2.95$ in the MIC  resulted in $\hat{b}=1.07$). We integrated the magnitude distribution (\ref{tGR}) in discrete bins of width $0.1$ to conform to the rules of the  experiment. 

\subsection{Expected Number of Events}
\label{sec:prednum}

The expected number of events per year in each space-magnitude bin $\left(i_x,i_y,i_m \right)$ was calculated from
\begin{equation}
E\left(i_x,i_y,i_m \right)= \lambda \ \mu^*(i_x,i_y) \ P(i_m)
\label{eq:expEv}
\end{equation}
where $\mu^*$ is the normalized spatial background density; $P(i_m)$ the integrated probability of an earthquake in magnitude bin $(i_m)$ defined according to equation (\ref{tGR}) and $\lambda$ is the expected number of earthquakes over a five- or ten-year period. To estimate the expected number of earthquakes, we counted the total number of observed $m \geq 4.95$ earthquakes in each (non-declustered) catalog and divided by the duration to obtain the mean number of events $m \geq 4.95$ per year. For the MIC, we estimated $\lambda^{mic}=1.24$ per annum, while there are an average of $\lambda^{cpti}=1.72$ earthquakes $m \geq 4.95$ per year in the CPTI catalog. To obtain five- and ten-year forecasts, we simply multiplied $\lambda$ by the number of years. Thus, based on the shorter MIC, we expect $6.2$ ($12.4$) earthquakes from 1 January 2010 until 31 December 2014 (until 31 December 2019), while based on the longer CPTI catalog, we predict $8.6$ ($17.2$) earthquakes over the same periods. 

\section{Five- and Ten-Year $m\geq 4.95$ Forecasts}
\label{sec:results}

The five-year forecasts based on the merged instrumental catalog (MIC) and the CPTI catalog are shown in Figures \ref{fig:m} and \ref{fig:c}, respectively. To obtain the ten-year forecasts, we doubled the rate in each space-magnitude bin because we assumed a temporally homogeneous Poisson process. The forecast based on the CPTI catalog is smoother than the map based on the shorter MIC because the 3,522 earthquakes of mostly small magnitudes $m \geq 2.95$ in the MIC cluster more strongly than the 623 events of larger magnitudes $m\geq 4.45$ in the CPTI catalog. The more evenly distributed epicenters of the CPTI catalog are more uncertain than those from the MIC, resulting in less clustering. 

We can compare the forecasts in Figures \ref{fig:m} and \ref{fig:c} with those of \citet[][Figure 4]{KaganJackson2010a} and \citet[][Figure 2]{ZecharJordan2010a}.  \citet[][]{KaganJackson2010a} used a fixed-bandwidth power-law kernel with an optimized bandwidth $r_s=5$ km to smooth seismicity in Italy listed in the PDE catalog above a threshold $m_t=4.7$. By visual inspection, their forecast is similar to the forecast based on the CPTI catalog in Figure \ref{fig:c}, although their fixed bandwidth of $r_s=5$ km is much smaller than the average of our optimal adaptive bandwidths $\langle d_i \rangle = 34.1$ (see model $14^!$ inTable \ref{tab:cpti}). The optimal smoothing bandwidths are different because \citet[][]{KaganJackson2010a} optimized their bandwidth for the 2004-2006 target period and smoothed earthquakes from a different data source. Given the observed dependence of the optimal smoothing distance on the chosen target period, we should expect to see differences in the optimal bandwidths. 

\citet[][]{ZecharJordan2010a} smoothed the CSI, the CPTI and a merged (``hybrid") catalog with an optimized fixed-bandwidth Gaussian kernel. Again, different choices for the magnitude threshold and the learning data make a direct comparison difficult, except for the forecasts based on the CPTI catalog optimized for the 2002-2006 target period. Figure 1 of \citet[][]{ZecharJordan2010a} shows that the optimal smoothing lengthscale is $\sigma=75$ km, while we obtained a mean bandwidth of $\langle d_i \rangle \approx 95.6$ (model 1 in Table \ref{tab:cpti}), indicating broad agreement between the two methods. Neither \citet[][]{KaganJackson2010a}  nor \citet[][]{ZecharJordan2010a} investigated whether the optimal smoothing length scale varies with target periods. 

\section{Discussion and Conclusions}
\label{sec:conc}

\citet{Werner-et-al2010d} evaluated all time-independent five- and ten-year forecasts of the CSEP-Italy experiment retrospectively on data from the CSI and CPTI catalogs. Using the forecasts from the first round of submissions from 1 August 2009, they found that several of the modelers had committed errors in the calibration of their models to calculate forecasts, principally in the conversion of the moment magnitude scale of the CPTI catalog to the local magnitude scale that will be used for prospective testing. Our first submission of the forecast based on the CPTI catalog also contained an error because of a mistake in the conversion formula we used (see equation \ref{eq:ML}). In this article, we only discussed the corrected forecasts submitted during the second round (1 January 2010). 

In the future, we would like to make a number of improvements to the model. First, we used a relatively arbitrary declustering procedure based on Reasenberg's algorithm \citep{Reasenberg1985}, although more objective methods exist \citep[e.g.][]{Zhuang-et-al2002, Console-et-al2010}. Second, contrary to the work by \citet{Helmstetter-et-al2007} and \citet{Werner-et-al2009b}, we could not estimate the completeness threshold as a function of space using their method and then attempt to correct for missing small earthquakes because the results were unstable. In the future, we would use a more robust method for estimating the completeness threshold. Third, we found that the optimal smoothing parameter varied substantially for different target periods, much more so than observed by \citet{Werner-et-al2009b}. The small number of target earthquakes might have caused the fluctuations. In the future, the optimal smoothing parameter should be optimized jointly over many target periods. More generally, we'd like to assess the influence of the choice of the kernel function. For example, do anisotropic kernels \citep[e.g.,][]{KaganJackson1994} improve the spatial forecasts? Does the optimal kernel choice depend on tectonic regime \citep[e.g.,][]{KaganJackson2010a, KaganJackson2009}? Should large earthquakes count more towards the density than small earthquakes \citep[e.g.,][]{Kagan-et-al2007}? 

Smoothed seismicity models make the implicit assumption that the available earthquake catalogs are long enough to estimate predictive spatial densities. \citet{KaganJackson1994}, however, conjectured that the optimal forecast horizon of an earthquake forecast based on smoothed seismicity scales with the duration of the learning catalog. To begin to address this question, we provided two earthquake forecasts: one based on a relatively short (30 years) data set of lower magnitude threshold and the other based on a longer (100 years) catalog with higher magnitude threshold. If the conjecture by \citet{KaganJackson1994} is correct, the forecast based on the merged instrumental catalog (MIC) should perform better than the forecast based on the CPTI catalog over shorter periods, while the CPTI-based forecast should show more predictive skill at longer timescales. 

\citet{Helmstetter-et-al2007} and \citet{Werner-et-al2009b} had previously found evidence for the hypothesis that including the locations of small earthquakes improves the forecasts of future epicenters of $m \geq 4.95$ earthquakes in California \citep[see also][for perspectives on the importance of small quakes]{Hanks1992, Marsan2005, Helmstetter-et-al2005, SornetteWerner2005a, SornetteWerner2005b}. The results of this study do not provide conclusive evidence for or against this hypothesis, perhaps because of the fluctuations induced by the small number of target earthquakes. A more robust cross-validation method for the optimal smoothing parameter might address this outstanding issue. 

The CSEP-Italy experiment, like its predecessor RELM, requires the use of the Poisson distribution for the number of earthquakes per space-magnitude bin \citep{Schorlemmer-et-al2010, Schorlemmer-et-al2007}. In principle, however, each model should provide its own model-dependent uncertainty bounds \citep{WernerSornette2008a}. For the time-independent Poisson process model described here, the distribution of the number of shocks over the relevant five- or ten-year timescales is assumed to be time-independent. However, the negative binomial distribution fits the number distribution better than the Poisson distribution \citep{Schorlemmer-et-al2010r, Kagan2010, Werner-et-al2009b, Werner-et-al2010d}. Therefore, in future iterations of the model, the Poisson distributions should be replaced by appropriate alternatives in each space-magnitude bin. A difficulty with this approach will be the estimation of parameter values based on small and possibly correlated samples. 

Nonetheless, despite the model's simplicity and its approximations, its five-year forecast submitted by \citet{Helmstetter-et-al2007} to the RELM experiment outperforms competitors after the first 2.5 years \citep{Schorlemmer-et-al2010r}. Whether the model can perform similarly in Italy, a different tectonic region from the seismically much more active California, will be an interesting test of the universal applicability of the model's assumptions. 

\section*{Data and Sharing Resources}

We used three earthquake catalogs for this study: the {\it Catalogo Parametrico dei Terremoti Italiani} (Parametric Catalog of Italian Earthquakes, CPTI08) \citep{CPTI08}, the {\it Catalogo del la Sismicit\`a Italiana} (Catalog of Italian Seismicity, CSI 1.1) \citep{Castello-et-al2007, Chiarabba-et-al2005}, and the {\it Bollettino Sismico Italiano} (Italian Seismic Bulletin, BSI) \citep{BSI2002, Amato-et-al2006}. The BSI is available at \url{http://bollettinosismico.rm.ingv.it}, and since July 2007 at \url{http://ISIDe.rm.ingv.it/}. The particular versions of the CSI and CPTI catalogs we used are available at \url{http://www.cseptesting.org/regions/italy}.

\section*{Acknowledgments}

We thank A. Christophersen, L. Gulia, J. Woessner and J. Zechar for stimulating discussions and F. Euchner for computational assistance. MJW was supported by the EXTREMES project of ETH's Competence Center Environment and Sustainability (CCES). AH was supported by the European Commission under grant TRIGS-043251, and by the French National Research Agency under grant  ASEISMIC. DDJ and YYK appreciate support from the National Science Foundation through grant EAR-0711515, as well as from the Southern California Earthquake Center (SCEC). MJW thanks SCEC for travel support. SCEC is funded by NSF Cooperative Agreement EAR-0106924 and USGS Cooperative Agreement 02HQAG0008. The SCEC contribution number for this paper is 1437.


\newpage

\section*{Tables}

\begin{table}[htbp]
  \centering
  \begin{tabular}{@{} |c|cccr|cccr|ccrr| @{}}
    \hline
    & \multicolumn{4}{c|}{Input Catalog (MIC)} & \multicolumn{4}{c|}{Target Catalog (MIC)}  &  \multicolumn{4}{c|}{Results} \\
    Model &  $t_1$ & $t_2$ & $m_{t}$ & $N_l$ & $t_1$ & $t_2$ & $m_{min}$ & $N_t$ & L &  G & $k$ & $\langle d_i \rangle$\\ 
    \hline
    1 &  1984 & 2003 & 2.95 & 2,632  & 2004 & 2008 & 4.95 & 6 & -45.4  & 2.12 & 5 &  14.2 \\ 
    2 & 1984 & 2003 & 3.45 & 755  & 2004 & 2008 & 4.95 & 6 & -45.0 & 2.26 & 4 &  25.5 \\ 
    3 &  1984 & 2003 & 3.95 & 223  & 2004 & 2008 & 4.95 & 6 & -44.5  & 2.44 & 2 & 32.5 \\ 
    4 &  1984 & 2003 & 4.45 & 63  & 2004 & 2008 & 4.95 & 6 & -44.1  & 2.60 & 1 &  48.8 \\ 
    5 &  1984 & 2003 & 4.95 &  19 & 2004 & 2008 & 4.95 & 6 & -45.5  & 2.07 & 1 & 95.6 \\ 
    6 &  1984 & 2003 & 5.45 &  5 & 2004 & 2008 & 4.95 & 6 & -48.1  & 1.34 & 1 &  221.3 \\ 
    \hline
    7*  & 1984 & 1998 & 2.95 &  1,900 & 1999 & 2003 & 4.95 & 8 & -65.4*  & 0.94* & 50*  & 60.1\\ 
    8  & 1984 & 1998 & 3.45 &  550 & 1999 & 2003 & 4.95 & 8 & -64.6  & 1.04 & 48  & 123.3 \\ 
    9 & 1984 & 2003 & 3.95 & 147  & 1999 & 2003 & 4.95 & 8 & -63.0  & 1.27 & 27  &  206.3 \\ 
    10  & 1984 & 2003 & 4.45 & 27  & 1999 & 2003 & 4.95 & 8 & -63.1  & 1.25 & 8 &  248.7 \\ 
        11  & 1984 & 2003 & 4.95 & 12  & 1999 & 2003 & 4.95 & 8 & -63.3  & 1.22 & 4  & 290.9\\ 
    12  & 1984 & 2003 & 5.45 & 4  & 1999 & 2003 & 4.95 & 8 & -63.7  & 1.16 & 2  & 500.9 \\ 
    \hline
    13  & 1984 & 1993 & 2.95 &  1,145 & 1994 & 1998 & 4.95 & 13 & -86.3  & 3.14 & 1  & 7.8 \\ 
    14 & 1984 & 1993 & 3.45 &  328 & 1994 & 1998 & 4.95 & 13 & -86.5  & 3.09 & 1  & 16.1\\ 
    15  & 1984 & 1993 & 3.95 &  81 & 1994 & 1998 & 4.95 & 13 & -99.4  & 1.15 & 5  & 107.5\\ 
    16  & 1984 & 1993 & 4.45 &  15 & 1994 & 1998 & 4.95 & 13 & -97.2  & 1.37 & 6  & 407.1 \\ 
    17  & 1984 & 1993 & 4.95 &  4 & 1994 & 1998 & 4.95 & 13 & -97.3  & 1.35 & 1  & 388.4\\ 
    18  & 1984 & 1993 & 5.45 &  0 & 1994 & 1998 & 4.95 & 13 & -  & - & -  & \\ 
    \hline
    \hline
    19  & 1984 & 2009 & 2.95 & 3,522  & 2004 & 2008 & 4.95 & 6 & -37.8  & 7.48 & 1  & 5.0 \\ 
    $20^!$ & 1984 & 2009 & 2.95 & 3,522  & 2004 & 2008 & 4.95 & 6 & -41.6$^{!}$ &   3.98$^{!}$ & $6^!$  & 13.8\\ 
    \hline
  \end{tabular}
  \caption{Results of the optimization of the spatial density estimate using the merged instrumental catalog (MIC) from 1 July 1984 through 25 June 2009. We varied the learning and target catalogs. The target catalog is the MIC in the testing region. The input catalog is the declustered MIC in the collection region. $N_l$  and $N_t$ are the number of earthquakes in the learning and testing catalog, respectively, $L$ is the log-likelihood score of a model,  $G$ is a model's probability gain per earthquake over a spatially uniform model, $k$ is the optimal number of neighbors to include in the bandwidth of the smoothing kernel, and $\langle d_i \rangle$ is the mean adaptive bandwidth in km. \newline
  $^{!}$ indicates that $k$ was not optimized but constrained to $k=6$. \newline
  $*$ denotes that the maximum $k=50$  of the range $[1,50]$ was attained.  }
  \label{tab:merged}
\end{table}

\begin{table}[htbp]
  \centering
  \begin{tabular}{@{} |c|cccr|cccr|crrr| @{}}
    \hline
    &  \multicolumn{4}{c|}{Input Catalog (CPTI)} & \multicolumn{4}{c|}{Target Catalog (CPTI)}  &  \multicolumn{4}{c|}{Results} \\
    Model  & $t_1$ & $t_2$ & $m_{t}$ & $N_l$ & $t_1$ & $t_2$ & $m_{min}$ & $N_t$ & L &  G & $k$ & $\langle d_i \rangle$ \\ 
    \hline
    1 &  1901 & 2001 & 4.45 & 605 & 2002 & 2006 & 4.95 & 7 &  -55.5 &  1.39 & 35 & 95.9 \\ 
    2 & 1901 & 2001 & 4.95 & 166 & 2002 & 2006 & 4.95 & 7 &  -54.8 &  1.54 & 12 & 102.9 \\ 
     3 &  1901 & 2001 & 5.45 &  51 & 2002 & 2006 & 4.95 & 7 &  -55.2 &  1.46 & 6 & 127.8\\ 
    \hline
    4 &  1901 & 1996 & 4.45 & 576 & 1997 & 2001 & 4.95 & 15 & -97.1   & 3.41 & 17 & 63.7\\ 
    5 &  1901 & 1996 & 4.95 & 155 & 1997 & 2001 & 4.95 & 15 & -101.8    & 2.49 & 8 & 84.6\\ 
    6 &  1901 & 1996 & 5.45 & 46 & 1997 & 2001 & 4.95 & 15 & -104.3   & 2.12 & 5 & 120.8 \\ 
    \hline
    7 &  1901 & 1991 & 4.45 & 565 & 1992 & 1996 & 4.95 & 2 & -15.9  & 4.41 & 2 & 17.8 \\ 
    8 &  1901 & 1991 & 4.95 & 153 & 1992 & 1996 & 4.95 & 2 & -16.5   & 3.17 & 1 & 22.7 \\ 
    9 & 1901 & 1991 & 5.45 & 46 & 1992 & 1996 & 4.95 & 2 & -18.0   & 1.53 & 8 & 199.3\\ 
    \hline
    10  & 1901 & 1986 & 4.45 & 551 & 1987 & 1991 & 4.95 & 6 & -46.6   & 1.72 & 2 & 18.0\\ 
    11 & 1901 & 1986 & 4.95 & 147 & 1987 & 1991 & 4.95 & 6 & -46.2   & 1.84 & 1 & 22.6 \\ 
    12 & 1901 & 1986 & 5.45 & 44 & 1987 & 1991 & 4.95 & 6 & -48.2   & 1.32 & 4 & 101.7\\ 
    \hline
    \hline
    13  & 1901 & 2006 & 4.45 & 623 & 2002 & 2006 & 4.95 & 7 &  -39.4    & 13.78 & 1 & 11.3\\ 
    14$^{!}$ & 1901 & 2006 & 4.45 & 623 & 2002 & 2006 & 4.95 & 7 &  -39.4$^{!}$    & 1.35$^{!}$ & 6$^{!}$ & 34.1\\ 
  \hline
  \end{tabular}
  \caption{Same as Table \ref{tab:merged} but using the CPTI catalog from 1901 to 2006 as data set. \newline
  $^{!}$ indicates that $k$ was not optimized but constrained. }

  \label{tab:cpti}
\end{table}

\clearpage


\begin{figure}
\centering
\caption{\label{fig:G} Probability gain per earthquake against magnitude threshold of the learning catalogs for various five-year target periods: blue -- merged instrumental catalog (MIC); red -- CPTI catalog; Cal. 2004-2008 -- the gains obtained for California by \citet{Werner-et-al2009b}; homogeneous -- the reference gain of a spatially homogeneous forecast. }
\end{figure}

\begin{figure}
\centering
\caption{Earthquake forecast based on the merged instrumental catalog (MIC): Expected number of earthquakes $m_L \geq 4.95$ over the five-year period from 1 January 2010  until 31 December 2014 per $(0.1^\circ)^2$ based on smoothing the locations of earthquakes $m_L\geq 2.95$ in the instrumental catalog from 1 July 1984 until 25 June 2009.}
\label{fig:m}
\end{figure}

\begin{figure}
\centering
\caption{Earthquake forecast based on the CPTI catalog: Expected number of earthquakes $m_L \geq 4.95$ over the five-year period from 1 January 2010 until 31 December 2014 per $(0.1^\circ)^2$ based on smoothing the locations of earthquakes $m_L \geq 4.45$ in the CPTI catalog from 1901 until 2006.}
\label{fig:c}
\end{figure}


\clearpage

\begin{figure}
\centering
\includegraphics[draft=\IsDraft,width=\halfwidth,keepaspectratio=true,clip]{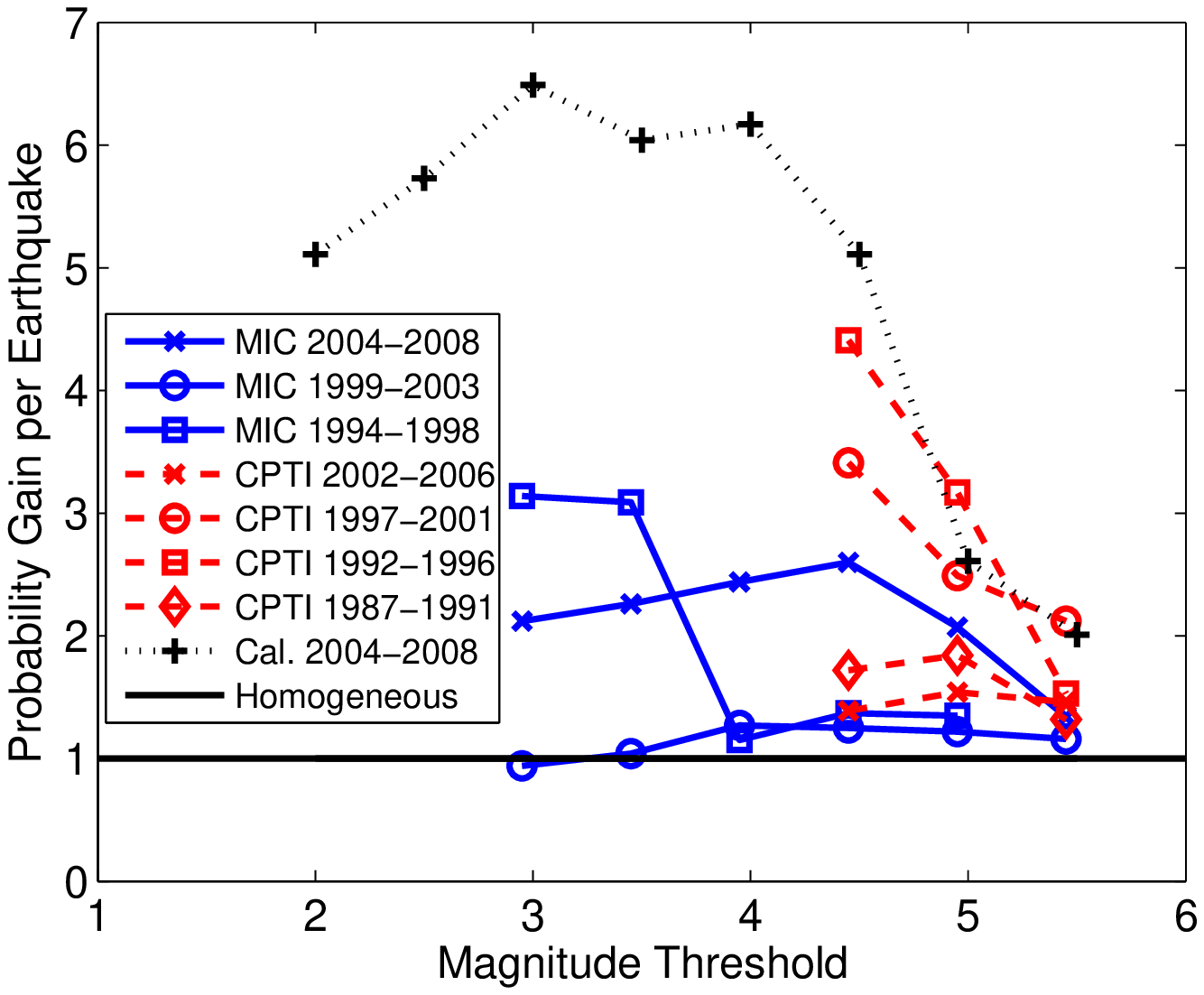}
\end{figure}

\begin{figure}[ht]
\begin{minipage}[b]{0.5\linewidth}
\centering
\includegraphics[draft=\IsDraft,width=8cm,keepaspectratio=true,clip]{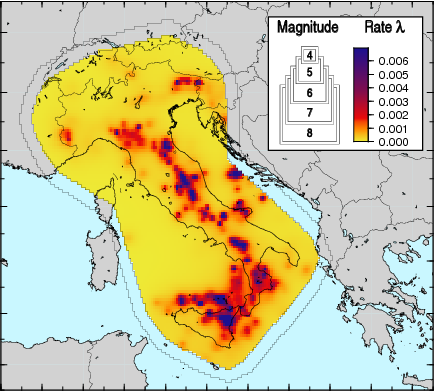}
\end{minipage}
\hspace{0.5cm}
\begin{minipage}[b]{0.5\linewidth}
\centering
\includegraphics[draft=\IsDraft,width=8cm,keepaspectratio=true,clip]{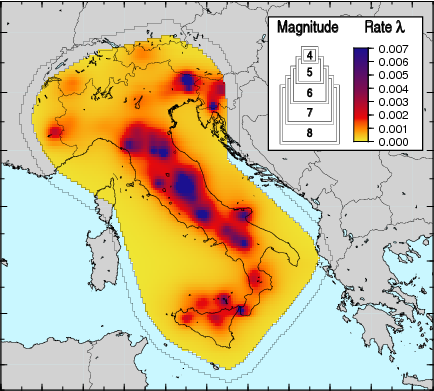}
\end{minipage}
\end{figure}

\end{document}